\newcommand{\ergcm}{\,erg\,s$^{-1}$\,cm$^{-2}$}
\newcommand{\ergs}{\,erg\,s$^{-1}$}
\renewcommand{\deg}{\ensuremath{^\circ}}
\newcommand{\phigh}{\ensuremath{P_{\rm{1.4}}}}
\newcommand{\shigh}{\ensuremath{\sigma_{\rm{1.4}}}}

\newcommand{\meanpcav}{\ensuremath{\bar{P}_{\rm{jet}}}}

\documentclass{emulateapj}

\usepackage{times,common}

\shorttitle{Mean jet power of 400SD clusters}
\shortauthors{Ma et al.}

\begin{document}

\title{Average Heating Rate of Hot Atmospheres in Distant Clusters by Radio AGN: Evidence for Continuous AGN Heating}

\author{
C.-J. Ma\altaffilmark{1,2},
B. R. McNamara\altaffilmark{1,2,3},
P. E. J. Nulsen\altaffilmark{2},
R. Schaffer\altaffilmark{1},
A.~Vikhlinin\altaffilmark{2,4},
}
\altaffiltext{1}{Department of Physics \& Astronomy, University of
  Waterloo, 200 University Ave. W., Waterloo, Ontario, N2L 3G1,
  Canada.}
\altaffiltext{2}{Harvard-Smithsonian Center for Astrophysics, 60
  Garden St., Cambridge, Massachusetts, 02138-1516, United States.}
\altaffiltext{3}{Perimeter Institute for Theoretical Physics, 31
  Caroline St. N., Waterloo, Ontario, N2L 2Y5, Canada.}
\altaffiltext{4}{Space Research Institute (IKI), Profsoyuznaya 84/32,
                 Moscow, Russia}

\begin{abstract}

We examine atmospheric heating by radio active galactic nuclei (AGN)
in distant X-ray clusters by cross correlating clusters selected from
the 400 Square Degree (400SD) X-ray Cluster survey with radio sources
in the NRAO VLA Sky Survey.  Roughly 30\% of the clusters show radio
emission above a flux threshold of 3 mJy within a projected radius of
250 kpc.  The radio emission is presumably associated with the
brightest cluster galaxy.  The mechanical jet power for each radio
source was determined using scaling relations between radio power and
cavity (mechanical) power determined for nearby clusters, groups, and
galaxies with hot atmospheres containing X-ray cavities.  The average
jet power of the central radio AGN is approximately $2\times
10^{44}$\ergs. We find no significant correlation between radio power,
hence mechanical jet power, and the X-ray luminosities of clusters in
the redshift range 0.1 -- 0.6.  This implies that the mechanical
heating rate per particle is higher in lower mass, lower X-ray
luminosity clusters.  The jet power averaged over the sample
corresponds to an atmospheric heating of approximately 0.2 keV per
particle within R$_{500}$.  Assuming the current AGN heating rate does
not evolve but remains constant to redshifts of 2, the heating
rate per particle would rise by a factor of two.  We find that the
energy injected from radio AGN contribute substantially to the excess
entropy in hot atmospheres needed to break self-similarity in cluster
scaling relations.  The detection frequency of radio AGN is
inconsistent with the presence of strong cooling flows in 400SD
clusters, but does not exclude weak cooling flows.  It is unclear
whether central AGN in 400SD clusters are maintained by feedback at
the base of a cooling flow.  Atmospheric heating by radio AGN may
retard the development of strong cooling flows at early epochs.

\end{abstract}


\section{Introduction}\label{sec:intro}

Heating by active galactic nuclei (AGN) may be responsible for two significant phenomena in  galaxy clusters. First, cool, dense cores found in the X-ray atmospheres of clusters  are expected to
cool efficiently to low temperatures \citep[][]{fabian94}.  However,  the levels of cold gas 
and star formation lie well below the values expected from pure
radiative cooling \citep[reviewed by][]{peterson06, mcnamara07}.  The
generally accepted solution is that the cooling is being compensated
by one or more heating mechanisms, possibly including conduction from
the hot gas in non-cooling regions \citep[e.g.][]{narayan01,
  zakamska03, voigt04, wise04}, and AGN heating
\citep[e.g.][]{binney95, churazov02, roychowdhury04, voit05}. Among
the possible heating mechanisms, AGN heating stands out because of
both the enormous energy released and the opportunity for feedback, since
AGN power is controlled by the accretion rate.  Second, galaxy groups
and poor clusters are less luminous than expected based on the $L_{\rm
  X}$-$T$ relation derived for massive clusters \citep{markevitch98, arnaud99, ruszkowski04, sanderson03,popesso05}. This is interpreted as a break of the self-similar scaling relation between X-ray luminosity
and gas temperature expected in structure formation models that include only gravitational heating \citep[][]{kaiser86,evrard96}.  The lower than expected X-ray luminosity for a given temperature is thought to reflect extra entropy (heat) in the gas that was injected by star formation and
AGN at early times \citep[e.g.][]{kaiser91, balogh99, borgani05, ponman03, croston05, sanderson05, jetha07, cavagnolo09}.   The ``preheating"  model proposes that the excess entropy was injected during the early stages of structure formation \citep[e.g.][]{kaiser91, balogh99, mccarthy02}. 

The amount of energy released by an AGN, on the order of
$10^{62}\,M_{\rm BH} / [{10^{9}\,{\rm M}_{\odot}}]$\,erg, is more
than enough to be the major heating source. Nevertheless, how and where the
energy is distributed into the ICM remains an issue. As discussed by
\citet{mcnamara07}, AGN heat in two ways.  First, AGN outbursts
inflate cavities.  The enthalpy released as the cavities rise
buoyantly is dispersed into the ICM \citep{bruggen02, reynolds02,
  churazov02, birzan04}.  Thermalization of the energy of cosmic rays
diffusing out of a radio lobe is an alternative channel for
converting lobe free energy to heat \citep[e.g.][]{mathews09}.
Second, the AGN inject energy into the ICM through shocks
\citep[e.g.][]{forman05, mcnamara05, nulsen07} and sound waves
\citep[e.g.][]{fabian05}.  Cavity power can be estimated using the
enthalpy contained in the cavity and its rise time. In practice, the
enthalpy is calculated as ${4pV_{\rm cav}}$, assuming the cavity is
filled with relativistic particles, and the age can be estimated using the terminal velocity of the buoyantly rising bubbles \citep[][]{birzan04,
  dunn05, rafferty06, birzan08}. All of these quantities, the pressure
of the surrounding gas $p$, the volume of the cavity $V_{\rm cav}$ and
sound speed in the ICM, can be obtained from X-ray data in principle. 
Simulations by \citep[][]{mendygral11} have verified that the cavity power estimated using this approach can be accurate to 
approximately a factor of two.  
Furthermore, \citet{birzan08} and \citet{cavagnolo10} find that the
radio power and the buoyant cavity power are correlated for samples
ranging from elliptical galaxies and galaxy groups to massive galaxy clusters. Thus, the radio power can be used as a proxy for the
cavity power for large samples of radio galaxies embedded in hot
atmospheres that lack deep X-ray observations. However, cavity power
estimated using this approach is probably a lower limit, because it
does not include the additional energy released by shocks
\citep[][]{forman05,mcnamara05,nulsen05a,nulsen05b}.

It is more challenging to calculate the total AGN energy deposited
over the life of a cluster.  Since the power output of AGN is highly
variable and outbursts are intermittent, current AGN power is not
necessarily representative of its historical average.  Even under the
simplest assumption, that the power of an AGN is the same in every
outburst \citep[which is inconsistent with data,
  e.g.][]{fabian06,randall11}, an estimate is still required for the
fraction of the time that the AGN is in outburst, i.e., for its duty
cycle, which is also inaccessible for individual AGN.  An alternative
to studying single systems is to determine the average AGN power for a
large sample.  For a well chosen sample, the average power should
represent the time average of the AGN power for individual sample
members.  The key to success here is to use a well-defined,
unbiased, and sufficiently large sample. 

In this paper, we estimate the average jet power in distant, X-ray
selected clusters found in the 400 Square Degree  Cluster Survey
\citep[400SD;][]{burenin07,vikhlinin05} by cross correlating the 400SD
catalog with the  NRAO VLA Sky Survey \citep[NVSS;][]{nvss}.  We
calculate jet powers for radio sources above a flux limit of
$\sim3$\,mJy.  Following earlier work on AGN heating in cooling core
clusters \citep[][]{dunn06,dunn08,rafferty06,rafferty08}, we evaluate
the AGN power injected into normal clusters to address the question of how much energy is deposited by AGN into
cluster halos over a significant fraction of their ages.  Note that
AGN feedback would be expected to enhance the incidence of radio
outbursts at the centers of cooling core clusters.  As a result, this
question cannot be answered based on a sample of nearby cooling core
clusters, which does not represent the general cluster population.
Also, since we are interested in the total power injected by AGN,
multiple radio sources may be attributed to a single cluster in the
present study.

The paper is arranged as follows. In \S\ref{sec:data}, we introduce
the 400SD clusters, and the matched radio sources. In particular, we
discuss the background source correction in
\S\ref{sec:data_nvssbackground}, and examine the morphology of a
subsample of the NVSS sources using images from the survey Faint
Images of the Radio Sky at Twenty-cm \citep[FIRST; ][]{first} in
\S\ref{sec:data_first}. Then we present our analysis and discuss the
interpretation: the correlation between the 1.4\,GHz power of the
radio AGNs and the X-ray luminosity of their host clusters in
\S\ref{sec:power_correlation}, the radial distribution of radio
sources in the clusters in \S\ref{sec:distribution_radio}, the jet
power estimated from the radio power in \S\ref{sec:cavitypower}, and
the estimation of AGN heating energy in \S\ref{sec:agnheating}. We summarize our results
in \S\ref{sec:discussion}.

\section{Sample}\label{sec:data}

\subsection{X-ray data}\label{sec:data_X-ray}

\begin{figure}[h] 
\hspace*{-5mm}\plotone{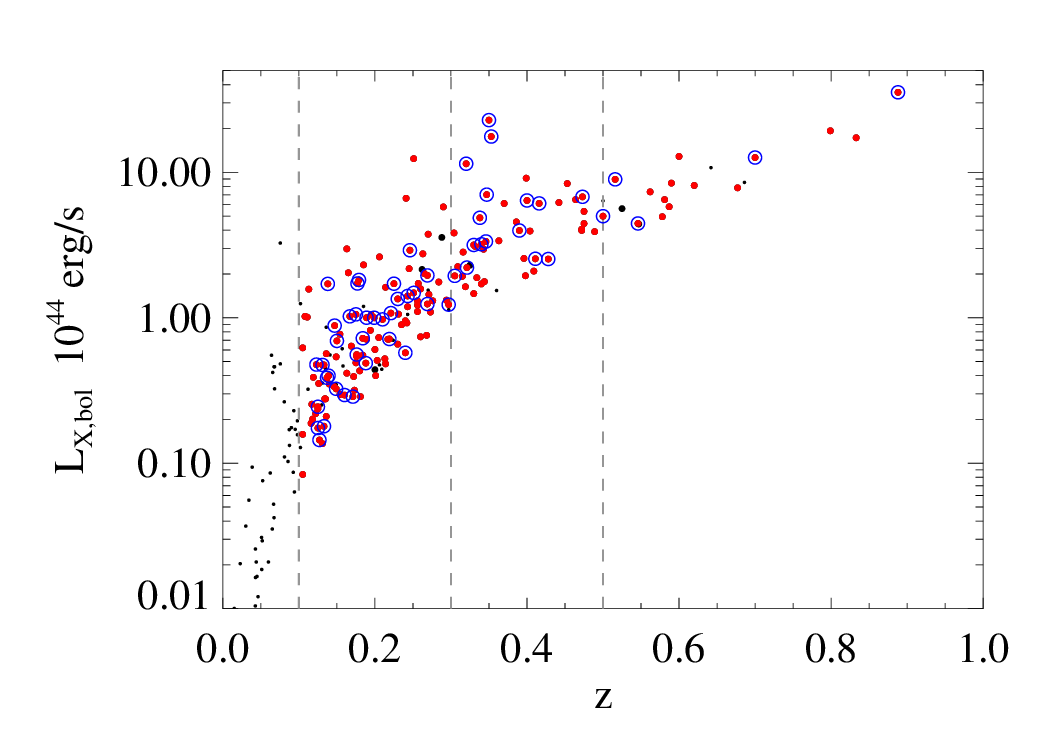}
\figcaption[]{X-ray bolometric luminosity distribution with redshift. Red dots mark the clusters with matched radio sources within $2$\,Mpc radius. Those clusters with radio sources at the center (r$<250$\,kpc) are plotted with blue circles. For the clusters located at the region of NVSS survey and no radio sources detected are  marked with black spheres. The other 400SD clusters with $\rm z<0.1$ are marked with  dots. The Y-axis is Bolometric luminosity derived using PSPC flux in $0.5-2.0$\,keV assuming L$_{\rm X}$-T relation (see \S\ref{sec:data_X-ray}). \label{fig:xlum}}
\end{figure}

The 400SD survey is one of the largest serendipitous X-ray surveys of clusters based on ROSAT Position Sensitive Proportional Counter pointed observations. It includes 242 optically confirmed clusters and groups brighter than an X-ray flux limit of $1.4\times10^{-13}$\,\ergcm\, in the $0.5-2$\,keV energy band, and lying at redshifts ${\rm z}< 0.9$  within a  $397$~deg$^2$ area. It is arguably one of the best  surveys for studying the evolution of clusters because the 400SD  samples ``normal'' clusters at higher redshift.  It is not targeted at the most massive clusters.  It includes clusters with L$_{\rm X,bol}\sim 10^{44}$~\ergs\ at ${\rm z}> 0.3$ (Fig.~\ref{fig:xlum}), in contrast to the all-sky surveys, such as MACS \citep{macs}, BCS+eBCS \citep{bcs_1, ebcs},
which are targeted at the rarer, more luminous and more massive clusters.  In addition, the ratio of strong cooling core clusters \citep[SCC; defined as cooling time less than $1$\,Gyr in][]{mittal09} in 400SD is very low according to \citet{santos10}, who find none SCC out of 20 clusters at $\rm z>0.5$), comparing to the all-sky surveys, e.g. BCS \citep{dunn08} and HIFLUGCS \citep{mittal09}. A similar conclusion is also made by \citet{samuele11} studying emission lines in the optical spectra of the brightest cluster galaxies (BCGs) in 400SD clusters.

Here we provide a short summary of the measurements of the 400SD clusters in the catalog of \citet{burenin07} and \citet{vikhlinin98b}.

\begin{itemize}
  \item The clusters were confirmed optically using images with magnitude limit m$_{R}\sim 24$. Their redshifts were measured spectroscopically.
  \item The center position and core radius of each cluster were derived from the best-fit $\beta$-model with $\beta$ fixed to $0.67$. The typical systematic error of the cluster positions is  $\sim 17\arcsec$ with respect to the brightest cluster galaxy.
  \item The X-ray flux in the energy band $0.5-2.0$\,keV was estimated as the mean of two fluxes calculated from the integration of best-fit $\beta$-models with $\beta$ fixed to $0.6$ and $0.7$ respectively. The $10\%$ systematic uncertainty caused by the uncertainty of $\beta$ value was added.  
  \item The X-ray Luminosity L$_{\rm X,0.5-2.0}$ was derived interactively assuming the L$_{\rm X}$-T relation of \citet{markevitch98} and \citet{fukazawa98}.
  \item To be consistent, we extrapolated to obtain the X-ray bolometric luminosity L$_{\rm X,bol}$ (Fig.~\ref{fig:xlum}) assuming the same L$_{\rm X}$-T relation. 
\end{itemize}  

\begin{figure}[h] 
\hspace*{-7mm}\plotone{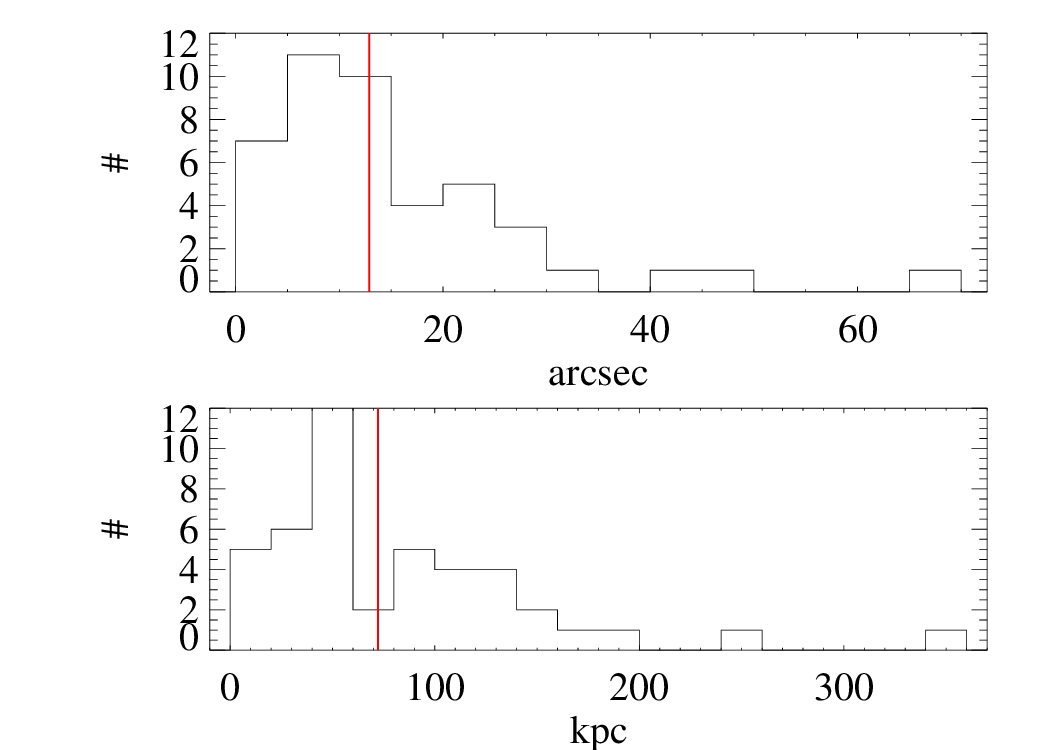}
\figcaption[]{Offsets between cluster centers measured using ACIS and
  PSPC data. The offsets are estimated from a subset (52) of clusters
  with ACIS images.  The upper histogram shows angular separation and
  the lower histogram shows projected distance at the cluster
  redshift.  The mean offset is marked by a red vertical line. \label{fig:coord_off}}
\end{figure}

We used the high resolution ACIS/Chandra data, which is available for
a subsample of the 400SD clusters \citep[][52 clusters in
  total]{vikhlinin09}, to examine the uncertainties in the luminosities
and centroids determined from the PSPC data.  Compared to centers
determined visually from the ACIS images (see Fig.~\ref{fig:coord_off}), we find that the average offset of the position in the catalog is $\sim 12\arcsec$, which is consistent with the uncertainty quoted above. In the rest frames of these clusters, all but one of the offsets correspond to projected distances of less than $\sim 250$\,kpc. The exception is a binary system (CL\,J0152.7-1357) whose cataloged centroid is located between the two clusters which are separated by $\sim100\arcsec$. 

In \citet{burenin07},  the uncertainty in the luminosities derived from the PSPC fluxes using more accurate Chandra fluxes is $19\pm0.6\%$. In addition, a few outliers, which are likely caused by contamination from X-ray point sources,  can be found in Fig.~23 of \citet{burenin07}.

\subsection{NVSS sources}\label{sec:data_nvss}

In order to match the coordinates of 400SD clusters with
1.4\,Ghz sources in the NVSS \citep[][]{nvss}, we extracted a
subsample consisting of the 400SD clusters in the region
  covered by the NVSS survey. This procedure yields 196 clusters at $\rm
z>0.1$. We find, in total, $782$ radio sources above a flux limit of
3\,mJy \footnote{The completeness reaches $90\%$ at 3\,mJy
  \citep{nvss}. We also tried a lower flux limit, 2\,mJy, and found
  that the results are qualitatively insensitive to the limit (see
  \S\ref{sec:agnheating}).}, located within a projected distance of
2\,Mpc of 166 clusters at $z>0.1$.  Of these sources, 61 are
located within a 250\,kpc radius of 56 clusters at $\rm
z>0.1$. Considering the uncertainty in the centroids for the 400SD
clusters (\S\ref{sec:data_X-ray}) and the resolution of NVSS images,
all sources within 250\,kpc are consistent with being associated
with the central galaxies of the clusters. We note that a small fraction of these matched
radio sources are background sources seen in projection and are not physically associated with the clusters. We discuss the background correction in
\S\ref{sec:data_nvssbackground}.

\begin{figure}[h] 
\hspace*{-3mm} \plotone{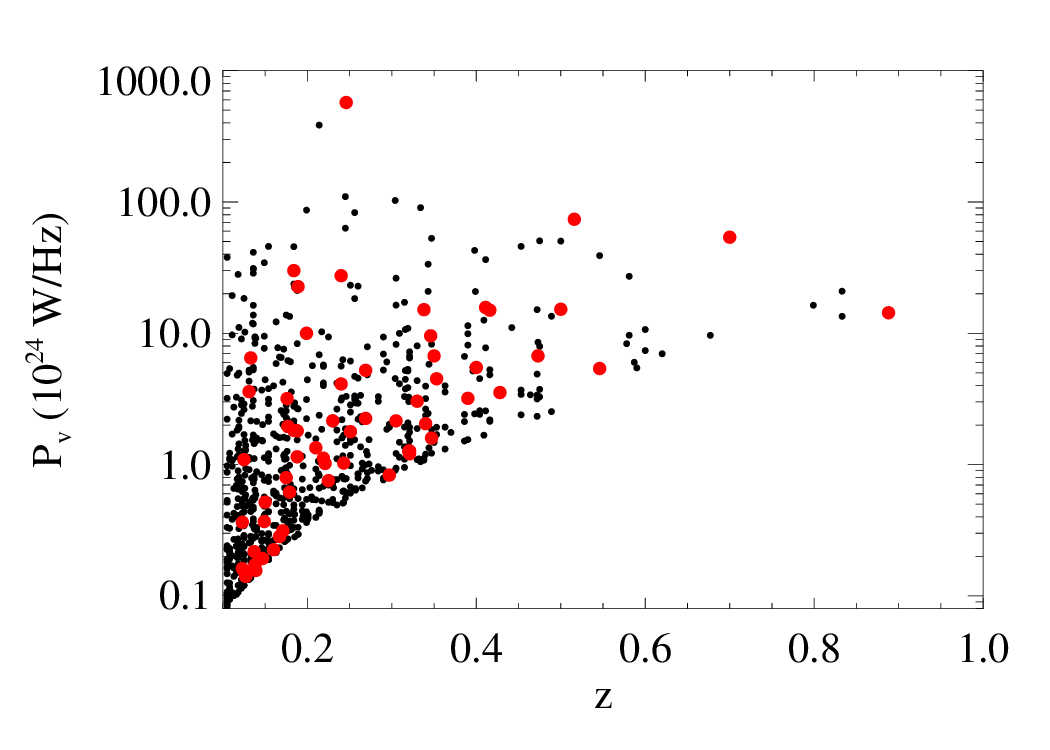}
\figcaption[]{Radio luminosity of NVSS sources around clusters. The luminosity density of matched NVSS sources are plotted with respect to the redshift of their associated cluster. The red spheres mark the sources within 250\,kpc radius of cluster core, and the smaller black dots mark those within 2\,Mpc radius. \label{fig:nvsslum}}
\end{figure}

\begin{figure}[h] 
\hspace*{-3mm}\plotone{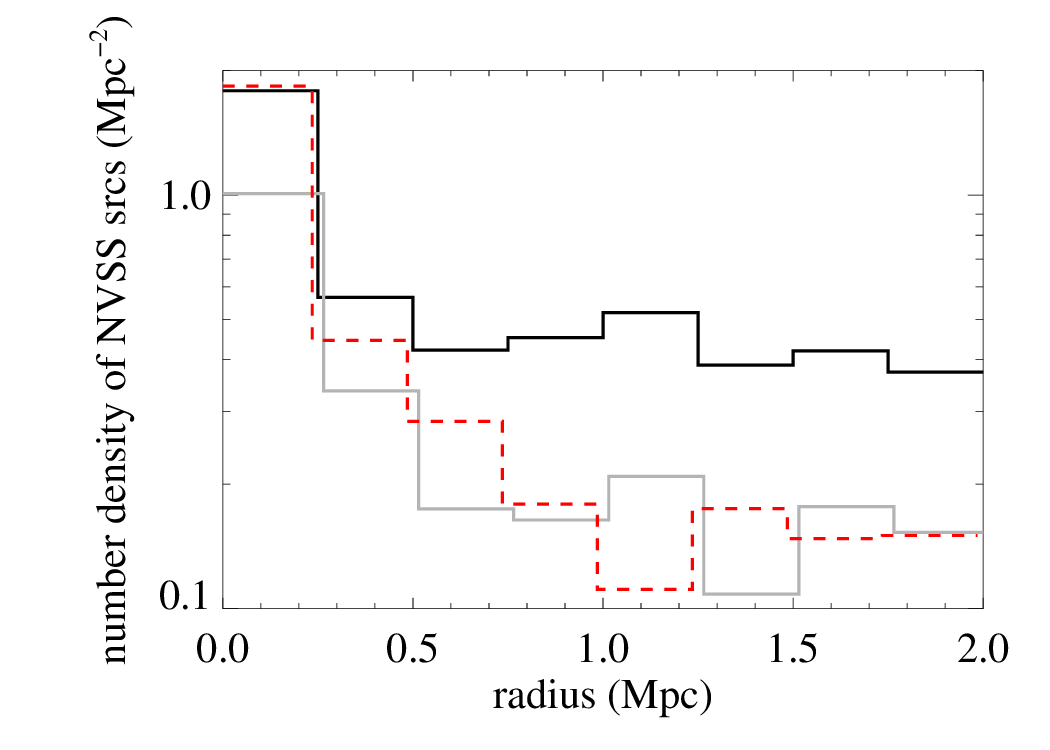}
\figcaption[]{Radial distribution of the surface density of NVSS
  sources around 400SD clusters. The NVSS sources are split into two
  redshift ranges: $0.1<{\rm z}< 0.3$ (black) and ${\rm z}> 0.3$ (red
  dashed) histograms. The gray histogram shows the surface density
  distribution for the low redshift sample with a luminosity cut
  P$_{1.4}>1\times10^{24}$~W\,Hz$^{-1}$. \label{fig:radialdistr}} 
\end{figure}
  
We calculated the radio power density, $P_{1.4}$
(Fig.~\ref{fig:nvsslum}), for each matched NVSS source, using the
cluster redshift and a spectral index of $-0.8$, which is the average for NVSS
sources \citep{nvss}.  The cluster redshift was used because we
cannot identify and measure the redshift of the host galaxy.

In Fig.~\ref{fig:radialdistr}, we plot the surface density of detected
NVSS sources around the 400SD clusters as functions of distance from
the X-ray centroid, for the redshift bins, ${\rm z}> 0.3$ and
$0.1<{\rm z}< 0.3$.  The surface density of NVSS sources associated with
clusters declines more quickly with distance at higher redshifts (red) compare to those at redshift between 0.1 and 0.3 (black).  The distribution for the lower redshift
sample is flat beyond $\sim 250$\,kpc.  The difference between the profiles arises because of the radio flux limit. Our adopted flux
limit of $3$\,mJy corresponds to a minimum luminosity of
$10^{24}\rm\ W\,Hz^{-1}$ (see  Fig.~\ref{fig:nvsslum}) for radio sources beyond ${\rm z}=0.3$, and drops to $10^{23}$\,W\,Hz$^{-1}$ for radio sources at the redshift $0.1$.    Thus, we detect many more low power radio sources within and projected onto the nearby clusters  compared to those at higher
redshifts.  Plotting the radial distribution of only the high luminosity sources
($P_{1.4} > 10^{24}\rm\ W\,Hz^{-1}$) for clusters lying between $0.1<{\rm z}< 0.3$ (gray
histogram in  Fig.~\ref{fig:radialdistr}) gives a closer match to
the distribution for clusters beyond ${\rm z}=0.3$.   However, the surface density of NVSS sources within
the central 250\,kpc is significantly higher in the higher redshift
clusters (red dashed histogram) compared to the lower redshift clusters (gray histogram). 

\subsection{Background correction}\label{sec:data_nvssbackground}

\begin{figure}[h] 
\plotone{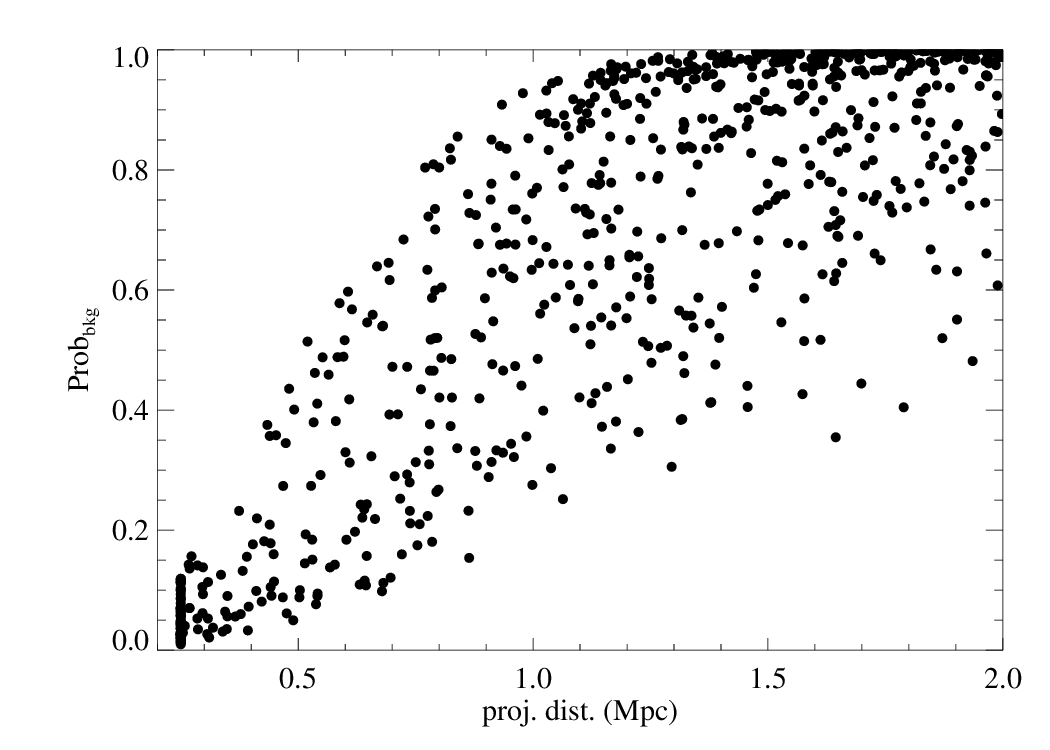}
\figcaption[]{Probability that a detected source is from the background.
  Projected distances for the sources located within $250$\,kpc are
  set to $250$\,kpc, reflecting the uncertainty in the cluster centers.  \label{fig:exp_bkgflux}}
\vspace*{5mm}
\end{figure}

\begin{figure}[h] 
\plotone{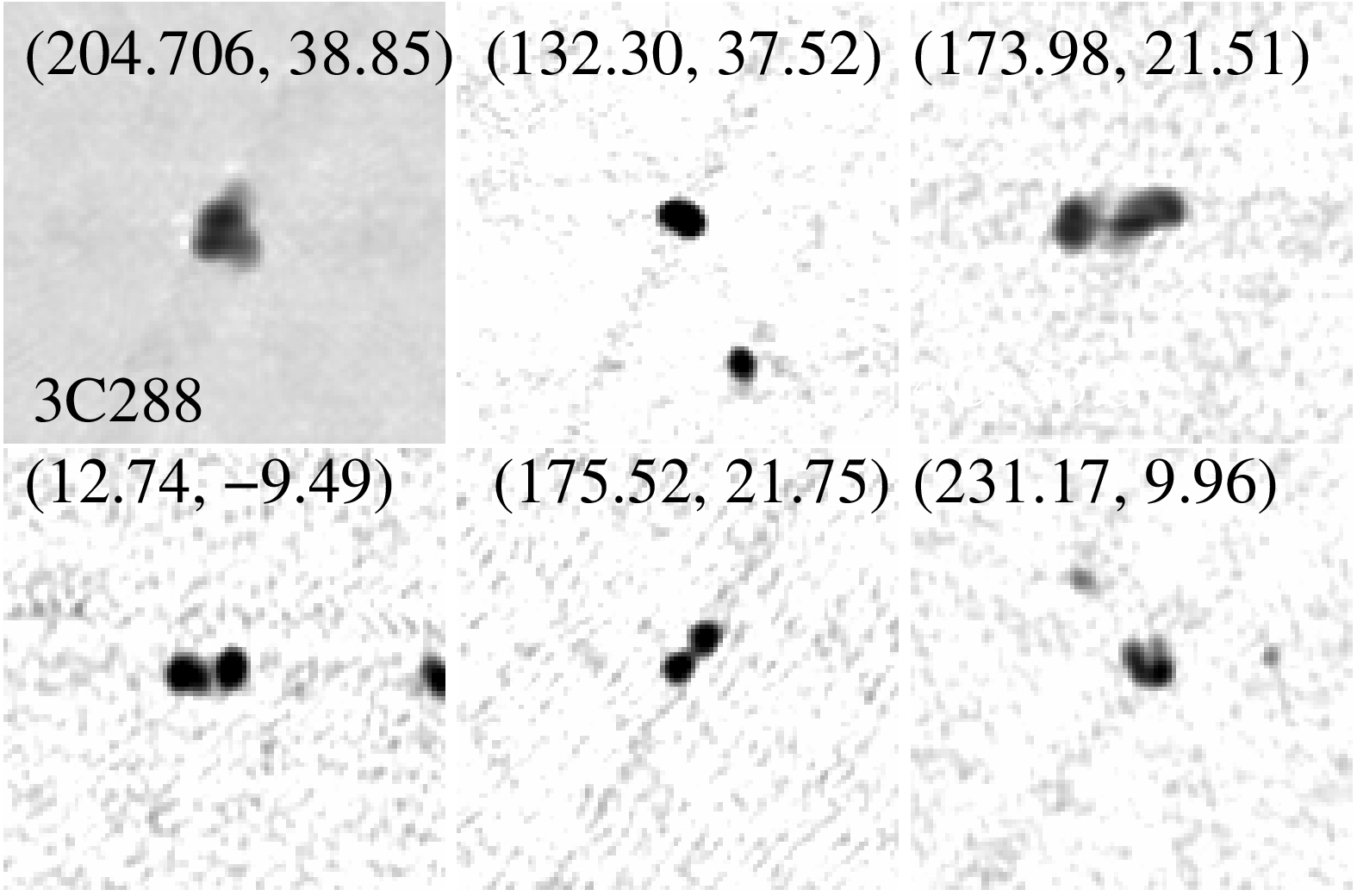}
\figcaption[]{FIRST images of matched NVSS sources.  We show the six
  strongest sources found within 250\,kpc, and the images (
  $3\arcm\times3\arcm$) are sorted by decreasingflux, from top left to
  bottom right.  The coordinates of the sources are
  noted on the top of each panel. \label{fig:examples}} 
\end{figure}
 
We are unable to identify the radio host galaxies due to the poor
resolution of the NVSS survey. Furthermore,  we do not have their
redshifts.  We have therefore corrected our numbers for background
contamination statistically. We assumed that the spatial distribution is locally homogeneous and
the flux function is independent to the location on the sky.  Thus,
the probability of finding any background sources within an area
$A$ can be expressed as 
\begin{equation}
P_{\rm bkg}(A) =1 - e^{-\lambda}, \qquad{\rm where} \qquad
\lambda = \frac{N_{\rm bkg}}{A_{\rm bkg}}\times A,
\label{eqn:pbg}
\end{equation} 
the mean $\lambda$ is estimated from the number density of the
background sources, the number of the NVSS sources in an annulus from 
$2\deg$  to $4\deg$ around the cluster of interest is  $N_{\rm
  bkg}$, and $A_{\rm bkg}$ is the area of this annulus\footnote{We have also tried the background source density calculated using the entire NVSS source catalog, and found no significant difference in our results.}. The
typical number of sources in the annulus is about 1600 and the
background number density is about 42~deg$^{-2}$, consistent with the
number density mentioned in \citet{nvss}.   Fig.~\ref{fig:exp_bkgflux} shows that any detected
source is likely to be a background source at large radii.  Therefore,
we confine out attention mostly to sources located within 250 kpc of
the center. 

In order to correct the flux contamination, we calculated the expected flux of background sources within the background area defined above. When we calculated the average flux of radio sources within some projected distance of a cluster,  we scaled the expected background flux according to the ratio of area, and subtracted it from the gross average flux. 
    
\subsection{High resolution radio source images for a subset of the sample}\label{sec:data_first}
   
Before we investigate the statistical properties of the matching radio
sources, we briefly discuss the properties of a few sources with
higher resolution radio images obtained from the FIRST data
archive. Overall, more than half of the NVSS sources were observed in
the FIRST survey. For example, we examined the images of 33 out of 56
NVSS sources matched within a 250\,kpc radius of the 400SD
clusters. We find 14 of these sources have double radio source morphologies,
and 7 of the remaining 19 single sources are resolved at the FIRST
resolution ($\sim 5\arcs$) with a minimum FIRST flux $\sim1.9$\,mJy.
We also checked the optical images of \citet{vikhlinin09}, and find
that $\sim 75\%$ (25/33) of the radio sources within 250\,kpc are
associated with BCGs.  We are unable to identify the BCGs of the remaining 8 clusters with radio sources using the available optical imagery.

In Fig.~\ref{fig:examples}, we show FIRST images for the strongest six
sources located within 250\,kpc radius of the 400SD clusters.  Of
particular interest is 3C288, which is the brightest NVSS source in
our sample (${\rm P_{1.4}}=700\times10^{24}$\,W\,Hz$^{-1}$). Its host is the BCG of cluster 400DJ1338+3851 at $\rm z=0.246$ with $\rm
L_{X,0.5-2.0}=9.6\times10^{43}$\,\ergs. It is a well-studied object
because of its bright but edge-darkened radio
morphology \citep{bridle89,ravi09,lal10}, and has been characterized
as a transitional FR\,I/FR\,II radio galaxy \citep{FR}. In the core of
the cluster, \citet{lal10} detected two surface brightness
discontinuities, which are interpreted as shocks induced by the
supersonic inflation of the radio lobes. The inferred  shock energy is
comparable to the cavity enthalpy of $\pcav=6.1\times10^{44}$\,\ergs.

\section{Results and Discussion}\label{sec:result}

\subsection{Correlation between radio power of AGN and X-ray luminosity of clusters?}\label{sec:power_correlation}

\begin{figure}[hb] 
\vspace*{-5mm}
\hspace*{-7mm}\plotone{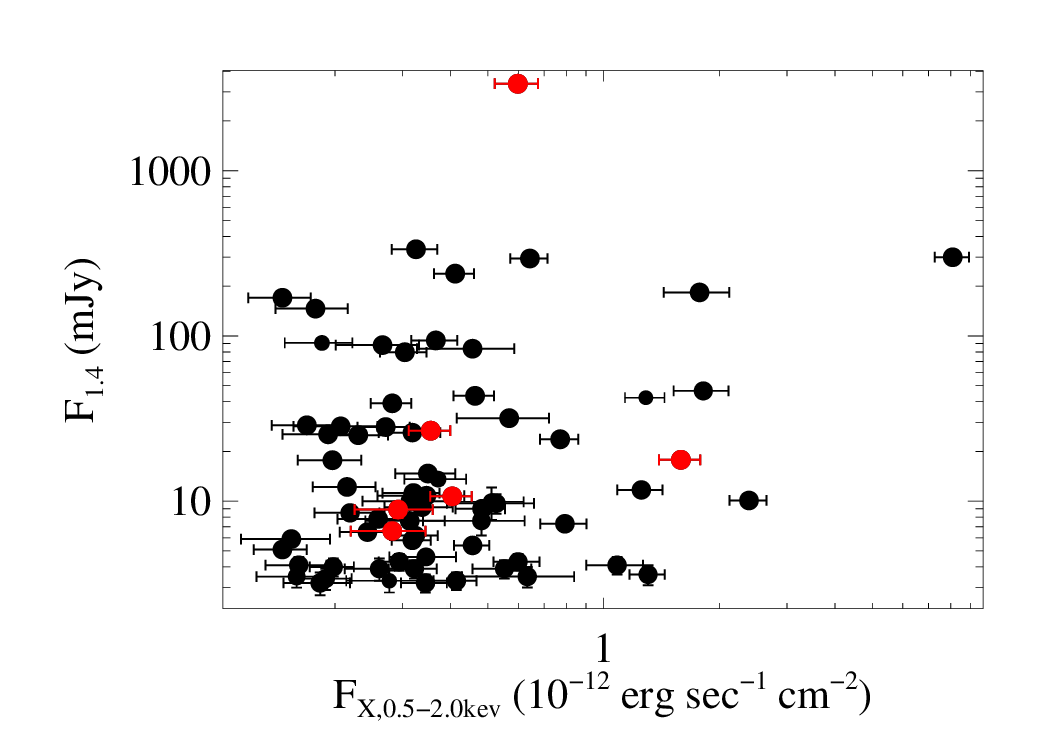}
\figcaption[]{$F_{1.4}$ versus $F_{\rm X}$ for the 400SD clusters with
  NVSS sources within $250$\,kpc.  The expected background fluxes
  (\S\ref{sec:data_nvssbackground}) have been subtracted, although the
  correction is insignificant for this small region. Using the archived Chandra data, we can estimate the cooling time of clusters referring from the concentration parameter $c_{SB}$. The clusters with cooling time less than 5\,Gyr are plotted in red. \label{fig:correlation_flux}}
\end{figure}

\begin{figure}[hb] 
\vspace*{-5mm}
\hspace*{-7mm}\plotone{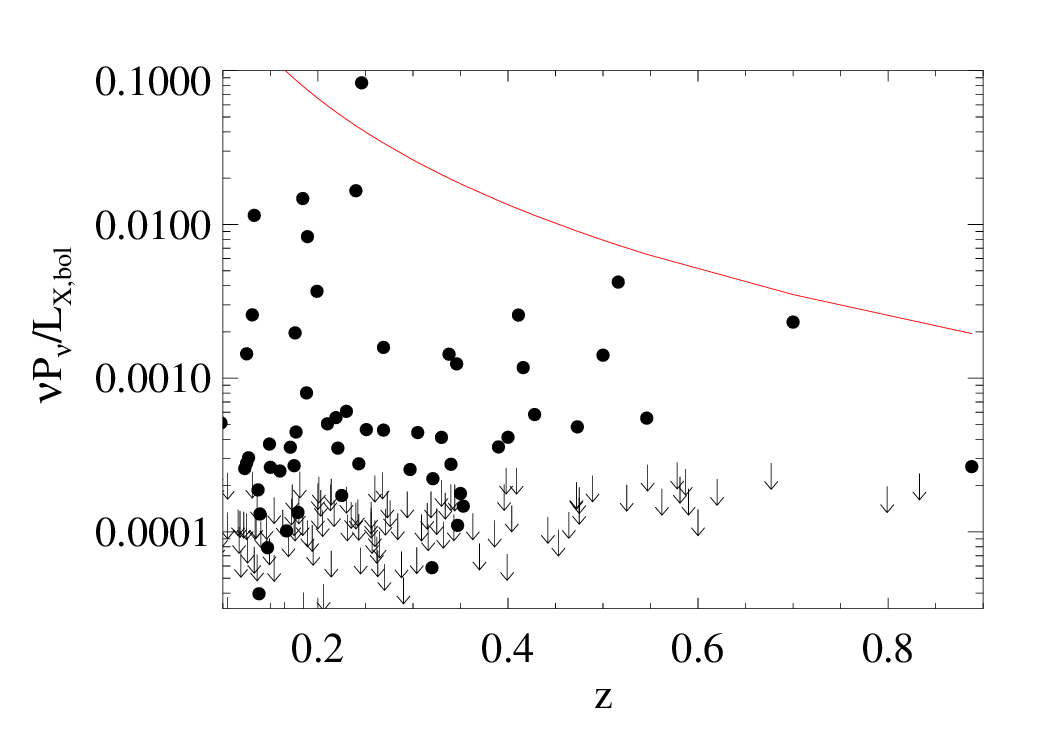}
\figcaption[]{Ratio of radio power within 250 kpc to X-ray luminosity
  for 400SD clusters.  Dots mark clusters with detected radio sources
  and upper limits are for non-detections, assuming one 3\,mJy radio
  source in each cluster.  The red curve shows the X-ray flux limit of
  the survey, for the median radio flux.  
\label{fig:correlation_powerratio}} 
\end{figure}

Using our catalog of 400SD clusters, we now examine the relationship
between the radio power of the central galaxies and cluster X-ray
luminosity.  A simple flux-flux plot in
Fig.~\ref{fig:correlation_flux} shows no correlation between the radio
fluxes\footnote{Radio fluxes are summed for all sources within 250\,kpc of each cluster cener. However, most of the clusters (51 out of 56) with radio sources within 250\,kpc radius only contain a single radio source, so that the fluxes could be taken as those of individual radio sources.} (F$_{1.4}$) of the sources within $250$\,kpc radius and the
X-ray fluxes (F$_{\rm X,0.5-2.0keV}$). The Spearman correlation is
only $0.15\pm0.20$.  To make sure the correlation is not hidden by
redshift dependent factors, such as reddening, we plot the ratio of
radio power to X-ray luminosity versus redshift in
Fig.~\ref{fig:correlation_powerratio}.  Although the points seem to be
centered on a constant ratio of $\sim 0.001$, the lower limits,
estimated for clusters with no detected NVSS sources
(F$_{1.4}>3$\,mJy), and the upper limit (red curve) for clusters not
detected at the X-ray flux limit of the 400SD survey, suggest that any
correlation is deceptive.  In summary, no correlation is found between
radio power and X-ray luminosity in our cluster sample.

Several previous studies \citep[e.g.][]{sun09,mittal09} have
demonstrated that the correlation between X-ray luminosity of the cooling core and the radio power only appears in ``strong" cooling core
clusters.  Therefore, the lack of a correlation between radio and
X-ray power is consistent with the lack of strong cooling core clusters
among 400SD clusters \citep{vikhlinin07,santos10,samuele11}. 
Of the 52 400SD clusters with Chandra data, 20 have NVSS sources within 250 kpc of the center.  Of those, 6 clusters have surface brightness concentrations $c_{\rm SB} > 0.1$ \citep[][]{santos08}, corresponding to central cooling times shorter than $\simeq 5$ Gy \citep[][]{santos10}.  These are likely to be weak cool core clusters \citep[WCC in ][]{mittal09}. These sources are plotted in red in Fig.~\ref{fig:correlation_flux} and their radio and X-ray fluxes do appear better correlated than for the bulk of the sample, consistent with previous studies.

\begin{figure}[h] 
\vspace*{-5mm}
\hspace*{-7mm}\plotone{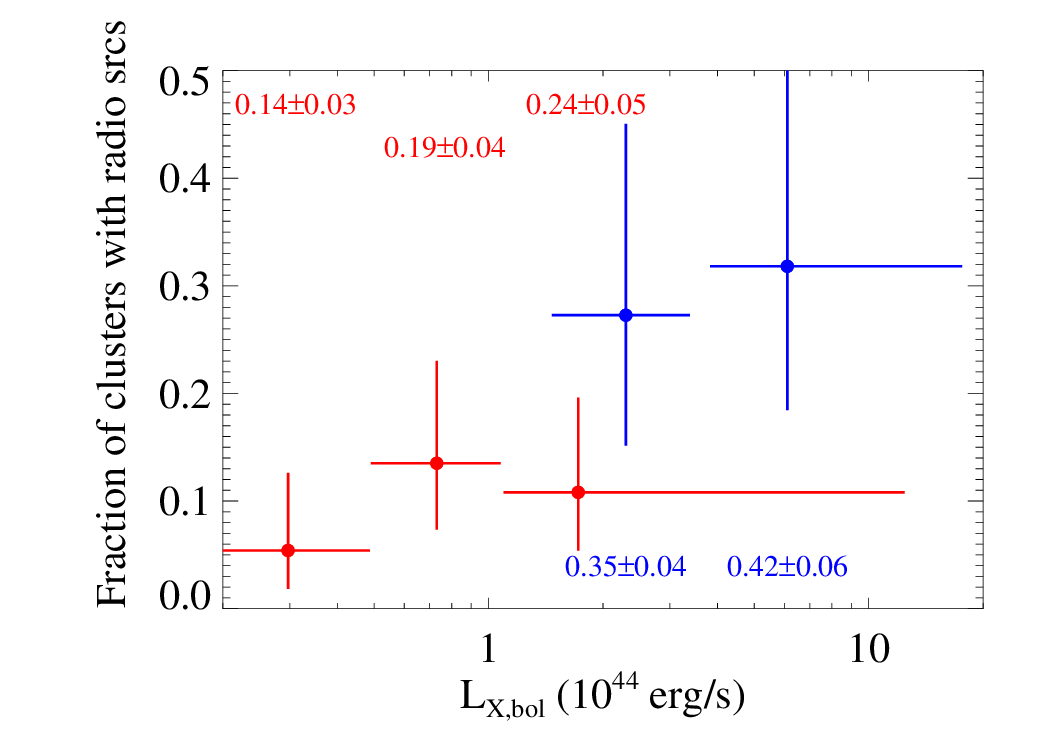}
\figcaption[]{Fraction of radio AGN in cluster cores versus X-ray
  luminosity.  The ratio of the number of clusters with radio loud AGN
  within 250\,kpc) to the total number of clusters in each X-ray
  luminosity bin is plotted against the X-ray luminosity.  To reveal
  the effect of the redshift-dependent luminosity cut, the clusters
  are separated into two subsamples by redshift, $0.3< z <0.5$ (blue)
  and $0.1< z <0.3$.  The mean redshift of each bin is marked on the
  plot.  Only radio sources with P$_{1.4}>3\,10^{24}$\,W\,Hz$^{-1}$
  are counted, corresponding to the radio power limit at ${\rm
    z}\sim0.5$.  Uncertainties are calculated assuming Poisson statistics. \label{fig:dutycycle}}
\end{figure}

We also consider the dependence of the fraction of clusters with NVSS
radio sources on X-ray luminosity.  
In Fig.~\ref{fig:dutycycle}, the incidence of radio sources appears to
increase slightly with X-ray luminosity.  However, the range of X-ray
luminosities sampled depends on the redshift in a flux-limited survey.
The low luminosity clusters in the sample are only detectable at lower
redshift because of the X-ray flux limit.  Separating the sample into
two redshift slices ($0.1< z <0.3$ and $0.3< z <0.5$ in
Fig.~\ref{fig:dutycycle}), we find that the apparent correlation
between X-ray luminosity and the fraction of cluster with NVSS sources disappears.  Thus, we suggest
that the difference between the blue points (higher redshift, more
luminous clusters) and the red points (lower redshift, less luminous
clusters) is probably caused by redshift and not X-ray luminosity.
\citet{best07a} found no correlation between the fraction of radio-loud
AGN in cluster galaxies and cluster velocity dispersion, which is, a
proxy for X-ray luminosity and cluster mass.  However, \citet{lin07}
found a clear increase in the ``radio active fraction" in a massive
cluster sample selected when they have bright BCGs with ${\rm
  M}_{k}<-24$.  In order to examine this apparent inconsistency, we
need to study the k-band luminosity of the BCGs which is beyond the
scope of this paper. 

Although we see no evidence of a dependence on X-ray luminosity, the
incidence of radio AGN does increase marginally at higher redshift.
This may reflect higher AGN fractions in the past, as suggested by
some, e.g.,by \cite{martini07, martini09} for X-ray AGN and by
\citet{galametz09} for AGN selected by radio, infrared and X-ray.
However, \citet{gralla10} analyzed radio AGN in the 618 clusters of
the Red Sequence Clusters Survey at redshifts $0.35< z <0.95$, finding
that the average number of AGN is independent of redshift.  Although
they are less significant, our results agree better with the trend found
for radio AGN by \citet{galametz09}, of an increase in incidence by
about a factor of 2 ($\sim 2\sigma$ significance) from $z<0.5$ to
$0.5<z<1.0$.

\begin{figure}[h] 
\vspace*{-5mm}
\hspace*{-7mm}\plotone{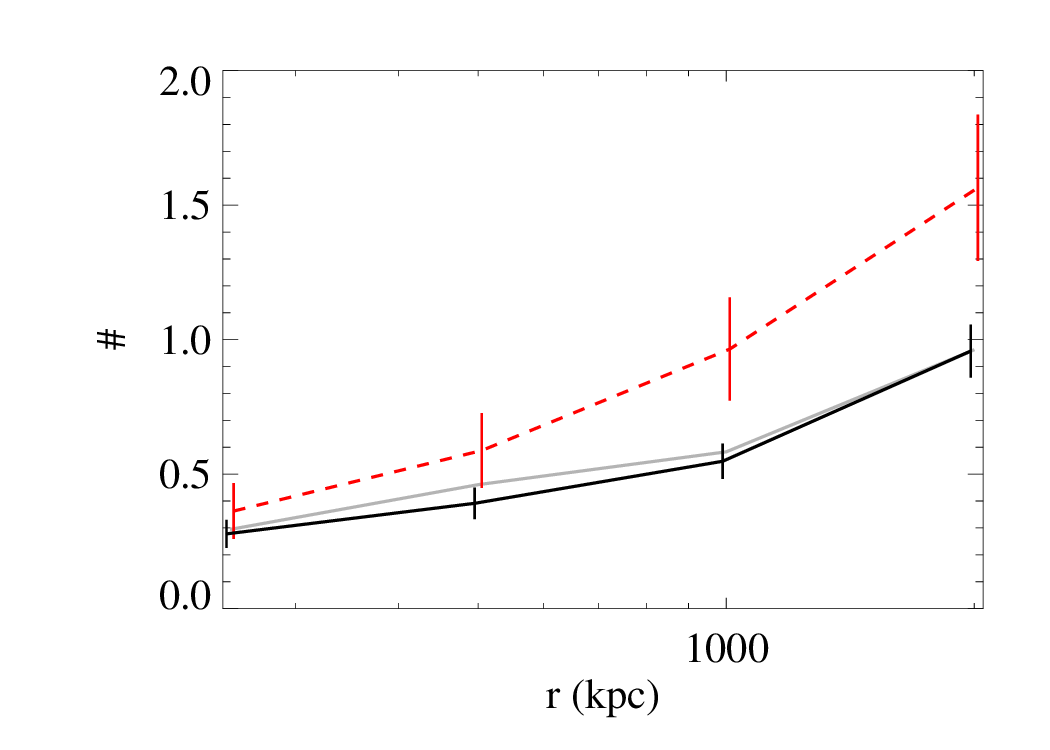}
\figcaption[]{Distribution of radio sources around 400SD clusters.
  The radio sources are grouped into two subsamples by redshift of the
  host cluster: $0.1< z < 0.3$ for the solid line and $0.3< z < 0.5$
  for the red dashed line.  The gray solid line shows the distribution
  for lower redshift sources, for clusters with $L_{\rm
    X,bol}>10^{44}$\,erg\,s$^{-1}$ for the host clusters.  The numbers
  have been corrected for background sources
  (\S\ref{sec:data_nvssbackground}).  Error bars show the standard
  deviations for each point, while those for the gray solid line are
  omitted for clarity.\label{fig:400sd_numsrcs}}
\vspace*{1cm}
\end{figure}

\subsection{Radial Distribution and Incidence of radio AGN in clusters}\label{sec:distribution_radio}

In Fig.~\ref{fig:400sd_numsrcs}, we plot the average cumulative number of radio
AGN within radius $r$ versus $r$ for the 400SD clusters.  The number
of radio sources increases from an average of $\sim0.3$ in the cluster
cores to $\sim 1$ within $2$\,Mpc.  However, the number of radio
sources increases much more slowly than the area (i.e., $\propto
r^2$), consistent with the observation that the surface density of
radio sources peaks at cluster centers \citep{ledlow95a, best07a,
  lin07, croft07}.

Since most 400SD clusters have no more than one radio source within 250 kpc,
the average number of radio sources shown as the leftmost point in
Fig.~\ref{fig:400sd_numsrcs} is equivalent to the radio source
fractions in Fig.~\ref{fig:dutycycle}.  The average number of radio
sources we find in the cores of 400SD clusters is significantly lower
than the fraction of radio-loud BCGs in nearby cooling core clusters,
which lies between 75\% to 100\% \citep{burns90,edwards07, dunn06b,mittal09}.
This fraction is expected to correlate with the central cooling time.
Our detection fraction is close to the fraction of radio sources found in nearby,
non-cooling core clusters ($45\%$) as defined by  \citet{mittal09},  and is identical to the fraction
found in optically selected clusters (30\%) determined by \citet{best07a}.
While a more quantitative comparison between the numbers is difficult,
a low radio source fraction for the 400SD clusters is consistent with
our claim that most of them lack strong cooling cores.  At the same time, our detection fraction
is significantly higher than the radio loud fraction of isolated ellipticals, which is $\sim 15\%$ \citet{best07a}.
So it is possible that our host galaxies lie at the centers of weak cooling flows.  This issue will need to
be addressed using deep X-ray imaging. 

In Fig.~\ref{fig:400sd_numsrcs}, the clusters are divided into two
redshift ranges, $0.1< z <0.3$ (solid line) and $0.3< z <0.5$ (red
dashed line).  We find that the average number of radio sources
increases with redshift, consistent with the discussion of
Fig.~\ref{fig:dutycycle} in \S\ref{sec:power_correlation}. Nevertheless, the comparison between the higher redshift and the lower redshift clusters   shows that the spatial distribution of radio sources around these clusters is not significantly different from that for all lower redshift clusters in the 400SD sample. 
 In order to show that the redshift dependence of the average number of radio sources is not a consequence of the X-ray
flux limit (i.e., luminosity limit) for the higher redshift clusters, we also plot the distribution of radio sources for lower redshift
clusters with $L_{\rm X, bol} > 10^{44}\rm\,erg\,s^{-1}$ (gray solid line). 
This X-ray luminosity corresponds approximately to the flux limit for the higher redshift sample, so the close correspondence between the gray and black curves shows that their difference from the radial distribution for the higher redshift sample (red dashed line) is not due to the X-ray flux limit.

\begin{figure}[h] 
\vspace*{-5mm}
\hspace*{-7mm}\plotone{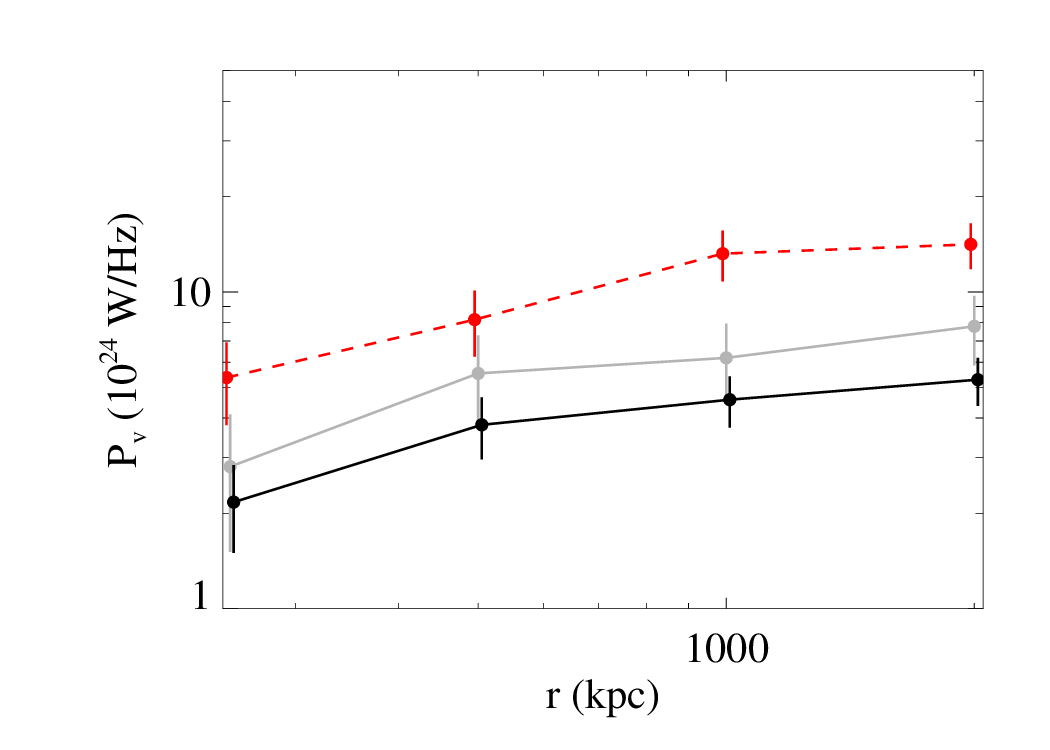}
\figcaption[]{Distribution of radio power around 400SD clusters.  The
  symbols are the same as for Fig.~\ref{fig:400sd_numsrcs}, but here
  the average power of NVSS radio sources within $r$ is plotted
  against $r$.   Note that the brightest source (3C288, with
  P$_{1.4}\sim 700\times10^{24}$\,W\,Hz\,$^{-1}$, see
  Fig.~\ref{fig:nvsslum}) at ${\rm z}\sim0.25$ is omitted (see
  \S\ref{sec:distribution_radio}).  If its radio power were included,
  the average powers for the lower redshift sources would be increased
  by $5\times10^{24}$\,W\,Hz$^{-1}$.  The uncertainty in the average
  power is calculated assuming a Gaussian distribution. \label{fig:400sd_accumpower}}
\end{figure}

Like Fig.~\ref{fig:400sd_numsrcs},
Fig.~\ref{fig:400sd_accumpower} shows the distribution of the average
radio power around 400SD clusters.  The radio power also show a modest
increase with redshift, but here the impact of the X-ray flux limit
(see the gray solid line) is more marked.   
Note that the average power for the low redshift sample does not
include the brightest radio source (3C288, ${\rm z}=0.246$), which is
roughly an order of magnitude brighter than the second brightest
source within $250$\,kpc (Fig.~\ref{fig:nvsslum}).  If the
contribution of this source was included, the average power of the low
redshift sample would exceed that for the high redshift sample.  We
omit this source because the jet power derived from its radio power
exceeds the jet power determined from its cavities by more than an
order of magnitude \citep[see \S\ref{sec:cavitypower}, ][]{lal10}. 
This discrepancy may simply reflect the large intrinsic scatter in the
relationship of Eqn.~\ref{eqn:high}, but we are unable to pursue this
issue here, since our sample lacks a statistically significant
population of such extremely luminous, short-lived sources.

\begin{figure}[h] 
\hspace*{-5mm}\plotone{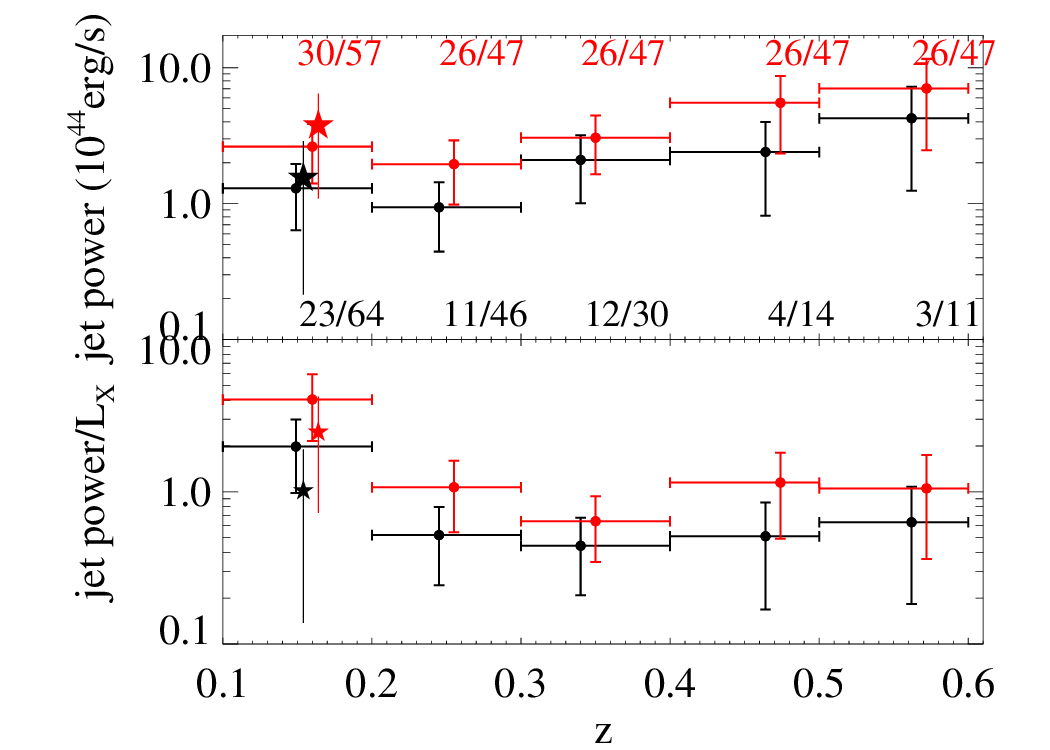}
\figcaption[]{Top: Average jet power (\meanpcav) of radio galaxies in the
  400SD clusters.  We estimate the jet power of radio sources from the
  1.4GHz flux in the NVSS catalogue, using the calibration of
  \citet{cavagnolo10}. The averages are calculated in 5 redshift bins
  covering $0.1< z < 0.6$, for the matched NVSS sources within
  $250$\,kpc (black circles) and 500\,kpc (red circles) of each cluster
  center.  For the lowest  redshift bin, results are also shown for
  subsamples with ${\rm L_{X}}>10^{44}$\,erg\,s$^{-1}$, plotted as
  stars.  The numbers of clusters hosting NVSS sources and the total
  numbers of clusters in each redshift bin are noted at the bottom/top
  of the plot.  The error bars, estimated using a Monte-Carlo method,
  are dominated by the scatter (\shigh) in the correlation between jet
  power and radio luminosity.  Bottom: Ratio of jet power to X-ray luminosity for 400SD
  clusters versus redshift.  The plot shows the average jet powers from
  the top panel of Fig.~\ref{fig:cavitypower} divided by the average total X-ray luminosity
  of the clusters in each redshift bin.  \label{fig:cavitypower}}
\end{figure}

\begin{figure}[h] 
\hspace*{-5mm}\plotone{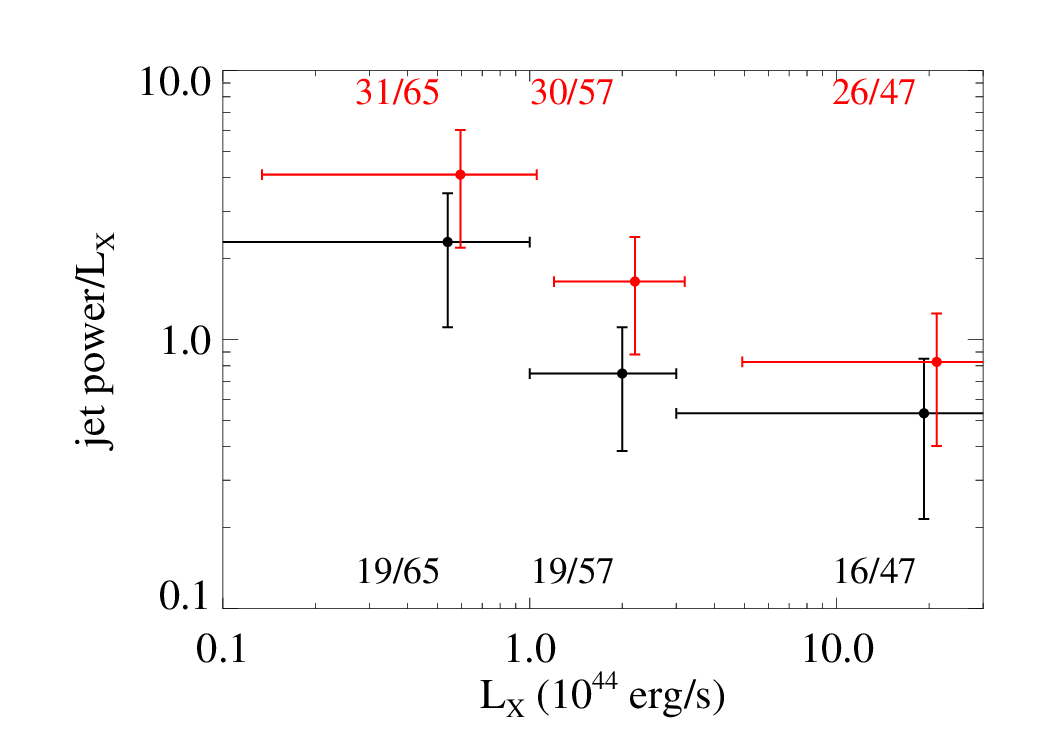}
\figcaption[]{Ratio of jet power to X-ray luminosity versus X-ray
  luminosity.  Similar to Fig.\ref{fig:cavitypower}, but here the
  sample is grouped by X-ray luminosity. \label{fig:cavitypower_lum_lum}}
\end{figure}

\subsection{Jet power estimated from radio power}\label{sec:cavitypower}

Radio power has been shown to correlate with ``cavity power,'' an
estimate of the mean power of radio jets based on enthalpies and ages
of X-ray cavities \citep{birzan04, birzan08, cavagnolo10}.  Given
their close correspondence to radio lobes, cavities are assumed to be
formed by AGN outbursts.  As a cavity rises buoyantly and expands, its
enthalpy is expected to be released into the hot atmosphere that hosts
it \citep[reviewed in][]{mcnamara07}.  This is one of several
mechanisms by which radio outbursts may heat the gas.  Other
mechanisms include shock waves or sound waves created by the expansion
of cavities \citep{forman05, mcnamara05,fabian05b, nulsen05a,
  nulsen05b, forman07} and thermalization of cosmic rays leaked from a
radio source \citep{bohringer88, mathews09,guo11}.  Here, we estimate jet
powers, \pcav, using the \phigh\ -- \pcav\ scaling relation of
\citet{cavagnolo10},
\begin{eqnarray}
  \log~\pcav\ = 0.75~(\pm 0.14)~\log~{\phigh\ } + 1.91~(\pm 0.18),  \label{eqn:high}
\end{eqnarray}
which is expected to provide, on average, a minimum estimate for the
total power injected into a cluster by a radio AGN.  Here, \pcav\ is
in units of $10^{42}\,{\rm erg\,s^{-1}}$, and \phigh\ in unit of
$10^{40}\,{\rm erg\,s^{-1}}$. The scatter in the correlation between
jet power and radio luminosity is $\shigh =0.78$\,dex.  It is
dominated by the intrinsic scatter \citep[][]{birzan08}, due to a
number of effects, including aging, and differences in particle
content and magnetic field.  The relationship (Eqn.~\ref{eqn:high}) is
determined over 6 decades in \phigh\, ($10^{37}$-$10^{43}$\,\ergs),
including data for nuclear radio sources for BCGs in cooling core
clusters, with $L_{\rm X}$ up to $10^{45}$\,\ergs, and low power
sources in giant ellipticals with $L_{\rm X}$ around $10^{43}$\,\ergs.
The large scatter in this scaling relation implies that the jet power
estimate it provides for any individual object has an uncertainty
approaching a factor of 10. 

We estimate the jet power for all of the cluster radio sources using Eqn.~\ref{eqn:high}, 
albeit with a large intrinsic uncertainty.  
Assuming the errors are random, we expect the uncertainty in the ensemble average 
to be reduced to an acceptable level.  More importantly, the average
jet power calculated for an unbiased sample should provide an estimate
of the contribution to the energy budget of AGN power integrated
across the sample.  A more precise measurement of jet power may be
obtained for an individual object by analyzing its X-ray cavities.  Despite being more precise, the measurement is only an instantaneous snap shot of its current jet power, and does not necessarily reflect the historical average AGN power.  
The ages of FR 1 radio AGN outbursts in clusters lie typically between $10^{7}$ to $10^{8}$\,yrs, and the lives of powerful FRII sources are even shorter \citep{odea09}. These time scales are much less than cluster ages, which greatly exceed $10^9$ years. Therefore, an estimate of the
time average jet power is required to calculate the total
AGN energy input to the clusters over cosmic time.  This quantity may be estimated by averaging jet power over a
sufficiently large and unbiased sample of clusters, as we have done here. 

In Fig.~\ref{fig:cavitypower}, we show our estimate of the average jet
power (\meanpcav), using a Monte-Carlo method that allows for the
uncertainty in the radio flux and the parameters in
Eqn.~\ref{eqn:high}, the distribution of radio spectral indices\footnote{A Gaussian distribution with a mean of 0.75 and a $\sigma$ of 0.10 \citep[][]{condon92} is assumed for the spectral indices.}, and
the large intrinsic scatter (\shigh) in the relation (Eqn.~\ref{eqn:high})
to estimate confidence ranges.  Overall, the confidence ranges shown
in Fig.~\ref{fig:cavitypower} are dominated by \shigh.  Note that we
use the arithmetic mean value of $\pcav$, excluding the brightest
source (see \S\ref{sec:distribution_radio}).  While this gives a mean
power that is substantially greater than the ``mean" of the log-normal
distribution of jet powers, assumed implicitly in the correlation
(Eqn.~\ref{eqn:high}), our purpose is to determine the average energy
deposited by AGN, not its distribution, so that the arithmetic mean is
the appropriate estimator.  In order to quantify the potential underestimate, 
we calculated the mean jet power excluding the 10\% most powerful radio sources, and found that the number drops by 5\%.

We find that the average jet power
in the cores of the 400SD clusters
is $2\times10^{44}$\ergs and does not show a significant redshift
evolution. The slight increase in the jet power at higher redshift
reflects the increased AGN fraction shown in Fig.~\ref{fig:dutycycle}
and Fig.~\ref{fig:400sd_numsrcs}.  The typical total X-ray luminosity
of the 400SD clusters (Fig.~\ref{fig:xlum}, $0.2<{\rm z}< 0.6$) is
$\sim3\times10^{44}$\,\ergs, so that the average jet power in these
clusters is about $50\%$ of their X-ray luminosity
(Fig.~\ref{fig:cavitypower}).  Comparing the two plots
in Fig.~\ref{fig:cavitypower}, we
find that the higher ratio of jet power to X-ray luminosity for the
lowest redshift bin shown in the bottom panel of Fig.~\ref{fig:cavitypower} is caused
primarily by selection effects.  Using an X-ray flux limit reduces the
average X-ray luminosity at lower redshifts, while lower luminosity
clusters at higher redshifts fall below the detection threshold.
Excluding low luminosity clusters ($L_{\rm X}<10^{44}$\ergs) for the
low redshift bin $0.1< z < 0.2$ gives an average jet power (star
symbols in Fig.~\ref{fig:cavitypower}) that is comparable to the
values at higher redshifts, consistent with the average radio power
being largely independent of the redshift, as argued in
\S\ref{sec:distribution_radio} (see Fig.~\ref{fig:400sd_accumpower}).
It suggests that the average jet power does not depend on the X-ray
luminosity of a cluster.  If so, the impact of AGN on the gas will be
greater in lower luminosity, hence less massive, systems.
Fig.~\ref{fig:cavitypower_lum_lum} shows that the minimum jet power
may exceed the power radiated by the ICM in group-scale objects with
$\rm L_{x}<10^{44}$\,\ergs.  If the jet power is mostly retained by
the hot atmospheres and not radiated away, AGN can cause a net
increase of entropy in low mass systems.


\subsection{AGN heating}\label{sec:agnheating}

\begin{deluxetable}{cc|cc|cc}
\tabletypesize{\scriptsize}
\tablewidth{0pc}
\tablecolumns{6} 
\tablecaption{Energy per particle\label{table:hpb}}
\tablehead{
\multicolumn{2}{c}{}                                                                       & \multicolumn{2}{c}{$<$250\,kpc}                                 & \multicolumn{2}{c}{$<$500\,kpc} \\
\colhead{N$_{\rm cl}$} & \colhead{${\bar{L}_{\rm X,bol}}$}  & \colhead{N$_{\rm det}$} & \colhead{heating}     & \colhead{N$_{\rm det}$} & \colhead{heating}     \\   
\colhead{}                        & \colhead{$10^{44}$\ergs}              &  \colhead{}                         & \colhead{keV\,par$^{-1}$}     & \colhead{}                          & \colhead{keV\,par$^{-1}$}  }
\startdata
65 & 0.47        &   19 &   $0.23\pm0.12$     & 56 &        $0.39\pm0.17$ \\
57 & $1.73$   &   19 &  $0.12\pm0.06$      & 42 &        $0.27\pm0.13$ \\
47 & $6.97$   &   16 &  $0.17\pm0.10$      & 36 &        $0.27\pm0.14$ \\
\cline{1-6} \vspace{-2.5mm} \\
169 & $2.71$ &  54  &  $0.17\pm0.08$     & 134&       $0.30\pm0.14$ 
\enddata
\end{deluxetable}

From the jet power, we can estimate the mean total energy per particle
injected by the radio sources (Table~\ref{table:hpb})
\begin{equation}
{\rm E}_{\rm jet}=\meanpcav\,t_{\rm z}{\mu}m_{\rm p}/\bar{M}_{\rm gas}, 
\end{equation}
where ${\mu}=0.59$ is the mean molecular weight and $m_{\rm p}$ is the
proton mass. $\meanpcav$ and $\bar{M}_{\rm gas}$ are the mean jet
power and gas mass, respectively, for the clusters\footnote{The mean jet power we estimated is defined within an aperture projected on the sky, but the gas mass here is defined in three dimension. From Fig.~\ref{fig:radialdistr} and discussions in \S\ref{sec:distribution_radio}, the radio sources in clusters are dominated by central sources. Therefore, the difference between projected and three dimensional mean jet power should not be significant, especially for small aperture like 250\,kpc.}.  The gas mass is
calculated from the X-ray luminosity assuming the $L_{\rm X}$-$M_{500}$ relation in \citet{vikhlinin09} and a gas mass fraction of $0.12$.  We excluded clusters beyond $z = 0.6$ because
their numbers are small.  The duration of energy injection, $t_z$, is
estimated as the time interval from $z=0.6$ to $0.1$ (4.4 billion years).  As might be
expected from the discussion of Fig.~\ref{fig:cavitypower_lum_lum},
${\rm E_{\rm jet}}$ decreases with X-ray luminosity, although the
trend is weaker due to the luminosity dependence of the $L_{\rm X}$-$M_{500}$ relation.  AGN heating from within the core alone
can reach about $0.2$\,keV/particle for poor clusters with typical
X-ray luminosities of $5\times10^{43}$\ergs.  The number increases to $\sim 0.4$\,keV/particle if the aperture increases to 500\,kpc. 
  We have also combined the data for the three
X-ray luminosity bins, giving the average AGN energy input per
particle for the entire sample in the last column of the table.

We expect the average energy injected by radio sources to be
underestimated somewhat due to the radio luminosity cut, particularly
for the high redshift clusters.  We examine this effect by calculating
the contribution from the weaker 
radio sources (${\rm P}_{1.4}<10^{24}$\,W/Hz) in clusters lying between redshift 0.1 and 0.3. We further assume that the radio luminosity function does not
evolve with redshift.  Including the contribution of the weaker
sources, the average energy injected by radio sources into
clusters at $0.3<{\rm z}<0.6$ is boosted by $7\%$ within 250\,kpc
radius and by $11\%$ within 500\,kpc.  This correction ignores radio
sources lying below our 3\,mJy radio flux limit.  Lowering the limit to $2$\,mJy and weighting to offset to the
flux-dependent completeness of the NVSS survey \citep{nvss}, the
increase in the average injected energy is negligible ($\le 2\%$).
Since the dependence of jet power on radio power according to
Cavagnolo's calibration is steeper than found previously by
\citet{birzan08}, the contribution of weak source is expected to be less important
than was found, for example, by \citet{hart09}.

We note that the numbers in the table are calculated using the mass of gas within R$_{500}$. The radio sources considered here deposit most of their energy directly into the cluster cores, but current ignorance of the significant heating processes means that we do not know where the bulk of the energy will ultimately reside.  If it did remain within the cluster cores, the energy deposited per unit mass would be approximately $\sim5$ times\footnote{The factor is estimated from the ratio of mass within  R$_{500}$ and 250\,kpc radius assuming a simple density profile $\rho \propto r^{-2}$.}) larger than the values given in the table, with a commensurate increase in its local impact. However, our primary objective is to demonstrate that the energy input is significant, even if the energy ends up being spread uniformly throughout much of the cluster atmospheres.

We have limited our jet power calculation in the redshift interval of the sample  ($0.1<\rm z<0.6$).
  In principle, the energy
injected by radio sources should be traced back to the time
when BCGs formed \citep[$z\sim2.0$, e.g.][]{vandokkum01}.  Ignoring
AGN evolution, the energy injected per particle since $z=2$ would be
at least twice the values listed in Table~\ref{table:hpb}.  This is a
conservative estimate, since the fraction of powerful AGN increases
with redshift \citep[e.g.][]{galametz09,martini09} and the
contribution of AGN heating is expected to be significantly greater in
the past.  Furthermore, we should take into account the evolution of
clusters over this time scale, so that we cannot simply scale the
values in the first three rows of Table~\ref{table:hpb}, which are
derived for fixed X-ray luminosity ranges.  Na\"ively, if we assume
that the mass function of 400SD clusters is representative of the
``integrated" mass function of clusters since $z=2$, we can
extrapolate the numbers in the last row of Table~\ref{table:hpb} to
provide a rough estimate of the minimum AGN energy injected per
particle since $z=2$. The values we find are $0.4$\,keV/particle from AGN within
250\,kpc.  

 \citet{wu00} found that the minimum excess energy required to break self-similarity is $\simeq 1$\,keV/particle.  Therefore, our results suggest that 
the continuous input of energy from AGN activity at 
the current rate over the ages of clusters would provide a significant fraction (40\%) of the excess
entropy (heat) found in the hot atmospheres of clusters.  It also implies that
AGN heating may be a factor leading to the dearth of strong cooling flows in distant 
clusters \citep[e.g.][]{vikhlinin07,samuele11}.  It is unclear whether this heating is
being supplied by a self-regulated feedback loop.

\section{Summary}\label{sec:discussion}

We have correlated the positions  of clusters from the 400SD cluster survey lying between redshifts 0.1 and 0.6  against radio sources  with  fluxes above 3\,mJy from the NVSS. Radio sources are detected within 250\,kpc of the center for 30\% of the clusters and within 1Mpc for 50 to 80\%, depending on redshift. The first value agrees with the fraction of radio sources in BCGs found by \citet{best07a} for SDSS clusters, but  lies well below the nearly 100\% detection rate in strong cooling flows \citep[e.g.][]{burns90,edwards07, dunn06b, mittal09, cavagnolo09}. We found that radio power of radio sources in the 400SD clusters does not correlate with the cluster X-ray luminosity. For comparison, \citet{mittal09} and \citet{sun09} found strong correlations between AGN radio power of BCG host galaxies and the X-ray luminosities
of their bright cooling cores, and the correlation disappeared in systems with weak or absent cooling cores.    Both of these facts support our conclusion that the 400SD sample is composed primarily of weak or non-cooling core clusters. 

We have found no significant correlation between the incidence of radio sources in clusters and the X-ray luminosity. Since the X-ray luminosity of a cluster correlates with its richness, this implies that richer clusters have fewer radio AGN per galaxy. In addition, the density profile of radio sources in clusters peak at the cluster center, implying that BCGs are the dominant radio galaxies in clusters. Apart from the BCG, the rich cluster environment is hostile to the formation of radio AGN, so that the number of radio AGNs does not increase with the number of galaxies. 

We have found only modest correlation between radio power and incidence of radio detection in 400SD clusters with redshift. Although it is difficult to completely rule out bias in a flux-limited sample, our results are qualitatively consistent with redshift evolution of the radio AGN detection rate in clusters by e.g. \citet{galametz09}.  

We have estimated the average jet power and the rate of AGN heating in clusters using the scaling relation of Eqn.~\ref{eqn:high}. The average jet power is about $2\times 10^{44}$\ergs for radio AGNs within 250\,kpc radius of the 400SD clusters and about $4\times 10^{44}$\ergs for sources within 500\,kpc radius. We found that the average jet power and AGN heating rate do not correlate with total X-ray luminosity of the 400SD clusters. Therefore, the heating rate per particle will be larger in less massive systems. The AGN heating within the core of clusters can reach 0.2\,keV/particle for poor cluster with typical X-ray luminosities of $5\times 10^{43}$\ergs. These numbers are calculated within the redshift range of our sample, i.e. $0.1<\rm z<0.6$. If we extrapolate the result to redshift $\rm z=2$ ignoring AGN evolution, the integrated AGN heating per particle will increase by a factor of 2. 

If the heated gas is unable to cool quickly enough, the entropy of their hot atmospheres  will rise above the values expected from gravitational heating alone.  In fact, we found in \S\ref{sec:agnheating} that the amount of AGN heating of the hot atmospheres is a significant fraction ($40\%$) of the heating required to ``preheat'' clusters \citep[][]{kaiser91,wu00,mccarthy02}.  This so-called preheating phase is thought to occur during the epoch of galaxy formation at redshifts of 3 and beyond \citep[][]{kaiser91}.  Our results suggest that the  AGN heating of cluster atmospheres long after the epochs of galaxy and cluster formation, and throughout the formation history of clusters are significant.  Thus, the heating that apparently broke the self-similarity of cluster scaling relations appears to have occurred continuously, and not necessarily at a single epoch.

\acknowledgements CJM and BRM are supported by Chandra Large Project Grant: G09-0140X. BRM acknowledge generous support from the Natural Sciences and Engineering Research Council of Canada.  PEJN was supported by NASA grant NAS8-03060.

\clearpage

\end{document}